\documentclass[%
 reprint,
 amsmath,amssymb,
 aps,
]{revtex4-2}

\usepackage{graphicx}
\usepackage{dcolumn}
\usepackage{bm}

\begin{document}

\preprint{APS/123-QED}

\title{Interplay of quantum confinement and strain effects in type I to type II transition in Ge/Si core-shell nanocrystals}

\author{Ivan Marri}
 \email{marri@unimore.it}

\author{Stefano Ossicini}%

\affiliation{%
 Department of Sciences and Methods for Engineering, University of Modena e Reggio Emilia, 42122 Reggio Emilia, Italy 
}%

\author{Michele Amato}
\affiliation{
 Universit\'e Paris-Saclay, CNRS, Laboratoire de Physique des Solides, 91405, Orsay, France
}%

\author{Simone Grillo and Olivia Pulci}
\affiliation{%
 Department  of  Physics,  University  of  Rome  Tor  Vergata, and INFN, Via  della  Ricerca  Scientifica  1,  I-00133  Rome,  Italy
}%

\date{\today}

\begin{abstract}
The electronic properties of hydrogenated, spherical,  Si/Ge and Ge/Si core-shell  nanocrystals with a diameter ranging from $1.8$ to $4.0$ nm are studied within Density Functional Theory. Effects induced by quantum confinement and strain on the near-band-edge states localization, as well as the band-offset properties between Si and Ge regions, are investigated in detail. On the one hand, we prove that Si(core)/Ge(shell) nanocrystals always show a type II band-offset alignment, with the HOMO mainly localized on the Ge shell region and the LUMO mainly localized on the Si core region. On the other hand,  our results point out that a type II offset cannot be observed in small (diameter  less than $3$ nm) Ge(core)/Si(shell) nanocrystals. In these systems, quantum confinement and strain drive the near-band-edge states to be mainly localized on Ge atoms inducing a type I alignment. In larger Ge(core)/Si(shell) nanocrystals, instead, the formation of a type II offset can be engineered by playing with both core and shell thickness. The conditions that favor the transition from a type I to a type II alignment for Ge(core)/Si(shell) nanocrystals are discussed in detail.
\end{abstract}

\maketitle


\section{Introduction}
\label{intro}
The electronic, transport and optical properties of silicon (Si) and germanium (Ge)  nanomaterials have been largely investigated in the past, both experimentally and theoretically, due to their promising applications in optoelectronics and photovoltaics \cite{Wolkin_PRL_99,Pavesi_Nature_2000,Melnikov_PhysRevB_2004,Nakamura_APL_2007,scarselli_APL_2007,govoni_nat,marri_JACS,Marri_Beilstein,Priolo_Nature_Tech_2014,Voros_J_Mat_Chem_A_2014,MARRI_SOLMAT,MARRI_pss,Kovalenko_ACS_Nano_2015,Degoli2009575,Angi_Nanoscale_2016,Dutta_ACS_Nano_2015,Sinelnikov_ACS_Phot_2017,FRAJ_ASS_2017,Delerue_PhysRevB_2018,Marri_PSSA_2018,TKALCEVIC_SEMSC_2020,Marri_PCCP_2020,Dhyani_IEEE_2020,Basioli_Appl_Nano_Mat_2020,Pavesi_Frontiers_2021,Marri_nanoscale_21}. Moreover, it has been shown that Si and Ge can be combined to obtain innovative materials that can be easily integrated into existing devices. Compared to pure Si and Ge materials, Si/Ge heterostructures offer more possibilities to tune the above-said properties  \cite{PaulSST2004,LeeJAP2005}. This can be achieved by varying Si and Ge atoms concentration and their spatial disposition, by modifying the geometry of Si/Ge interface and  by modulating both  strain and the quantum confinement effect (QCE) to obtain systems with the desired properties. \\
 \noindent
Si/Ge  heterostructures have been fabricated using different techniques, such as molecular beam epitaxy \cite{Ma_APL_2012}, self-assembly \cite{Lee_JES_2011,Alkhatib_SR_2013}, ion beam and magnetic sputtering deposition \cite{Buljan_PRB_2010,YANG_ASS_2012,Nekic_Nanophotonics_2017,Stavarache_SR_2020},  chemical vapor deposition \cite{BEI_APL_2007,Chang_APL_2010}, chemical synthesis \cite{Yang_Chem_Mat_1999} and gas-phase and nonthermal plasma synthesis \cite{Mehringer_Nanoscale_2015,HunterACS2017,Seraj_Eur_Phys_J_Appl_Phys_2020}. They have been integrated into different technological devices,  for instance in high-speed and high-power field-effect transistors \cite{Kenichi_APL_2005,Xiang_nature_2006,Liang_NL_2007,Mooney_Ann_Rep_Mat_Sci_2000, Rucker_Sem_Sci_Tech_2018}, photodetectors \cite{Shi_APL_2020,Palade_SEnsors_2020}, linear and non-linear optics devices \cite{Li_Nanotechnology_2020,Zhigunov_ACS_Photonics_2018}, solar cell systems  \cite{ZHAO_SEMSC_2017,LACHAUME_SEMSC_2017}, non-volatile memory  \cite{LiuIEEE2008} and thermoelectric  \cite{YamasakaSR2016} devices.\\
\noindent
Nowadays, the research focused on Si/Ge nanosystems, in particular core-shell (CS) nanowires (NWs) \cite{AmatoCR2014,Fukata_ACS_Nano_2015,Zhang_nanoscale_2018,Zhang_Nanoscale_2020} and nanocrystals (NCs) \cite{PiNanoT2009,BerbezierSSR2009,AquaPR2013},  represents one of the most rapidly developing areas in materials science.
CS nanosystems offer the possibility of  engineering electronic and optical properties by  varying core diameter and shell thickness  \cite{RamosPRB2005,JavanTSF2015,Cojocaru_SR_2021} (thus modulating  strain and QCE of both core and shell regions) and by  switching core and shell materials.\\
\noindent
Particular attention has been dedicated to the study of the band-offset properties of Si/Ge and Ge/Si  CSNCs, a fundamental step  to understand the relative localization of electrons and holes. The  alignment of energy levels of both core and  shell materials  can result in type I (band-edges localized on the same material) or in  type II (band-edges localized on different materials) heterostructures.
Type I materials show a strong overlap between electron and hole wavefunctions, that can be exploited in light-emitting devices. 
Type II alignment shows, instead, a weaker overlap between electron and hole wavefunctions which are, in this case, localized on different materials. This induces longer radiative lifetimes, lower excitation binding energies  and smaller exciton oscillator strengths when compared to type I structures.
Leading to a reduction of  non-radiative Auger recombination rates - because of the favored photogenerated charge carriers separation and their extraction - type II heterostructures are of great interest for photovoltaics applications \cite{TayakagiSR2013}. \\
\noindent
Energy levels alignment between Si and Ge bulks reveals a type II offset for Si/Ge superlattice  heterostructures, a band-edge profile that, however, can be altered  by strain and QC when low dimensionality is taken into account. 
 \\
\noindent
Band-offset properties of Si/Ge and Ge/Si CSNCs were extensively studied theoretically, with results not always consistent with each other.
A type II confinement, with electrons localized in the Si regions and holes localized in the Ge regions, was predicted for small Si/Ge and Ge/Si CSNCs by  Ramos et al. \cite{RamosPRB2005}. A similar result was obtained  by de Oliveira et al. \cite{OliveiraJPC2012}, through the analysis of the radial distribution of the HOMO and LUMO orbitals. However, while the formation of a type II offset  was clearly proved for the Si(core)/Ge(shell) NCs, it was less evident for the Ge(core)/Si(shell) NCs. The formation of a type II offset in small Ge(core)/Si(shell) NCs was  questioned by Nestoklon et al. \cite{NestoklonJPC2016}. By adopting a tight-binding model to study systems of different core and shell extensions, they always observed a strong localization of the LUMO state in the Ge core region for structures with a thin Si shell (thickness below  0.8 nm). 
Finally, a type II band-offset was predicted, using tight-binding methods, by Neupane et al. for large spherical Ge(core)/Si(shell) NCs with diameters ranging from 11 to 17.5 nm \cite{NeupaneJAP2011} and for large  dome-shaped GeSi CSNCs \cite{NeupaneJAP2012,Neupane_PCCP_2015}.\\
\noindent
Theoretical works agree therefore in indicating a type II offset for both Si(core)/Ge(shell) and large Ge(core)/Si(shell) NCs but do not uniquely define the band-offset properties of small Ge(core)/Si(shell) NCs. Moreover, an unambiguous description of the band alignment characteristics of Ge(core)/Si(shell) NCs as a function of the NCs size is still  missing.
In this work, we consider  Si(core)/Ge(shell) and  Ge(core)/Si(shell) NCs of different sizes and compositions and we clarify the mechanisms behind the formation of a type II offset, shedding light on the role played by the QCE and strain. To conduct our investigation, we adopt two different approaches. In the first one, the whole CSNC is taken into account without any simplification. This approach is used to obtain a quantitative signature of the band-offset. In the second method, core and shell are analyzed separately,  thus adopting an approach that allows to better clarify the role played by both QC (see Sect. \ref{SiGe_core_shell}) and strain (see Sect. \ref{strain}).


\section{Method}
The structural and electronic properties of different spherical,  Si(core)/Ge(shell) and Ge(core)/Si(shell) NCs, with diameters ranging from 1.8 to 4.0 nm, have been investigated by means of first-principles calculations. Only hydrogenated  NCs have been considered, thus  avoiding configurations leading to the formation of near-band-edge surface states. For all the considered systems,  we have performed Density Functional Theory (DFT) calculations using the Local Density Approximation (LDA) for the exchange-correlation functional, as implemented in the plane-wave pseudopotential Quantum ESPRESSO (QE)\cite{giannozzi_2009,Giannozzi_2017} code.
A careful analysis of the convergence of both structural and  electronic properties, in terms of plane-wave basis set cutoff, has been conducted. Norm-conserving pseudopotentials with a kinetic cutoff for the plane-wave basis set of 40 Ry have been adopted for all the considered systems. 
NCs have been placed in large cubic  cells containing a large amount of vacuum  to avoid spurious interactions between replicas. All atomic positions in the supercells have been fully relaxed until the forces acting on each atom were less than 0.003 Ry/a.u.  Si(core)/Ge(shell) and Ge(core)/Si(shell) band-edge properties have been determined by analyzing the   localization of Kohn-Sham (KS) HOMO and LUMO states, as done in previous studies (see for example Ref.~\cite{Nduwimana2008}). For one of the smallest NCs (Ge$_{35}$Si$_{112}$H$_{100}$), the band offsets were also estimated  within the GW approximation, to  check if the DFT-KS band ordering could be affected by many-body effects. Calculations have been performed 
adopting  80000 plane waves for the calculation of the exchange part of the self-energy $\Sigma_x$ and 30000 plane waves for both the screening $\epsilon^{-1}$ and the correlation part of the self-energy $\Sigma_c$. The  total number of bands was set to  2500 for $\epsilon^{-1}$ and to 3700 for $\Sigma_c$ \cite{chisig}.  
A spherical cut of the Coulomb interaction was used to avoid spurious interactions between periodic replicas. Noticeably, obtained results point out that DFT outcomes are not altered by the inclusion of GW corrections. More details are reported in  Sect. \ref{appendix}.

\section{Results}
In the following, we focus our attention on the study of the band-offset properties of Si(core)/Ge(shell) and Ge(core)/Si(shell) NCs of different sizes. The section is divided into four parts. In Sect. \ref{Alignment}, we discuss the intrinsic properties of Si and Ge materials. In Sect. \ref{CSNCs}, we analyse the band-offset properties of Si/Ge and Ge/Si CSNCs with diameters ranging from $1.8$ to $3.0$ nm. In Sect.   \ref{SiGe_core_shell} we compare the effects induced by the confinement of the electronic charge density  on the  electronic properties of systems of different shapes but with the same number of atoms. Finally, in Sect. \ref{strain}, we investigate the role played by strain. 
\begin{figure*}[t!]
    \centering
    \includegraphics[width=0.9\textwidth]{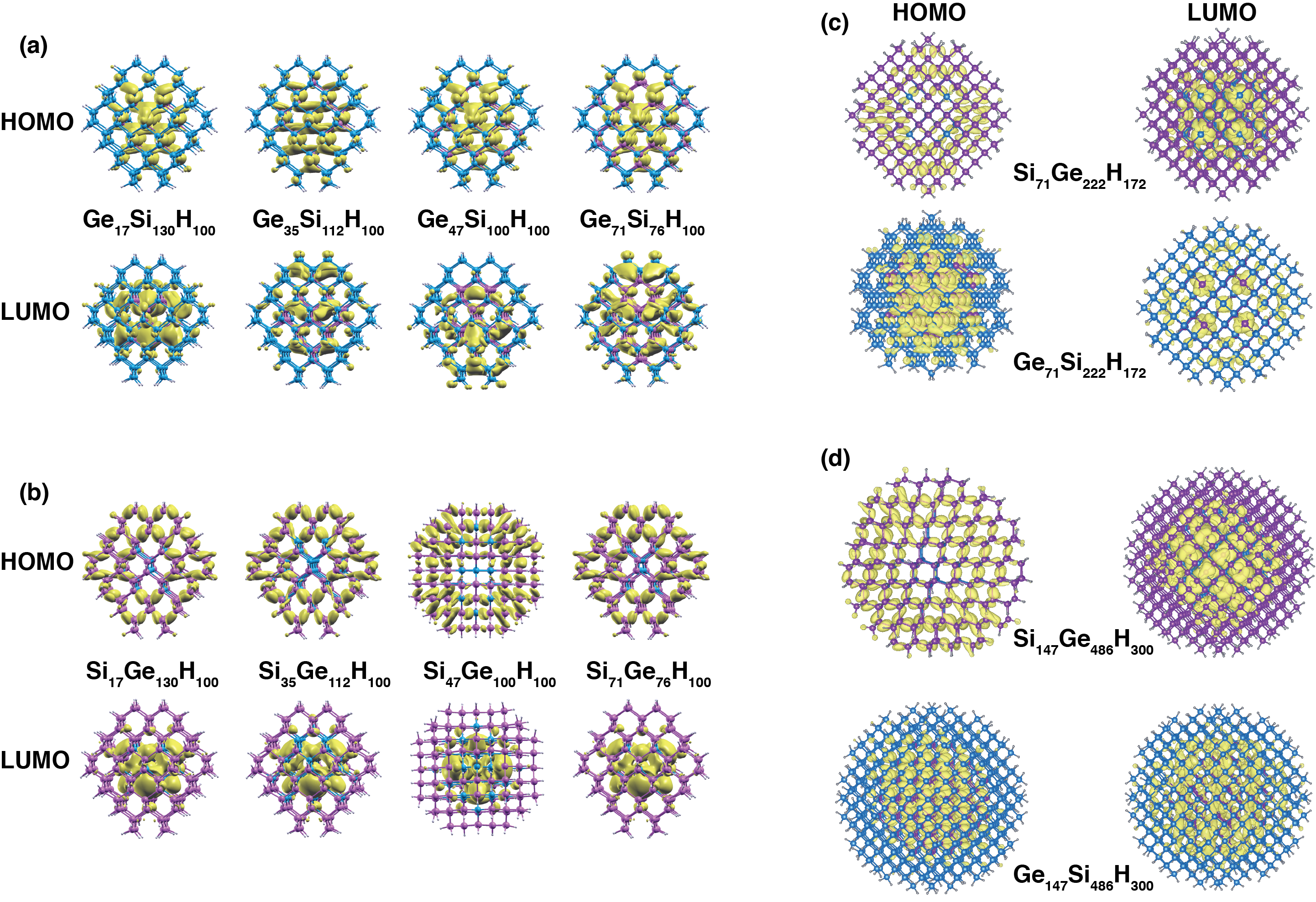}
    \caption{(Color online)  The spatial localization of the electronic wave functions of the HOMO and LUMO states for the  Ge$_{17}$Si$_{130}$H$_{100}$, the Ge$_{35}$Si$_{122}$H$_{100}$, the Ge$_{47}$Si$_{100}$H$_{100}$, the Ge$_{71}$Si$_{76}$H$_{100}$,  the Si$_{17}$Ge$_{130}$H$_{100}$, the Si$_{35}$Ge$_{122}$H$_{100}$, the Si$_{47}$Ge$_{100}$H$_{100}$ and finally  the Si$_{71}$Ge$_{76}$H$_{100}$ are reported on the left, panels (a) and (b), respectively. The parameter $P=N_{shell}/N_{core}$ ranges from $7.65$ to about $1$. On the right, the localization of the HOMO and LUMO wavefunctions  is depicted for the Si$_{71}$Ge$_{222}$H$_{172}$ (top row of panel c), the Ge$_{71}$Si$_{222}$H$_{172}$ (bottom row of panel c), the Si$_{147}$Ge$_{486}$H$_{300}$ (top row of panel d) and the Ge$_{147}$Si$_{486}$H$_{300}$ (bottom row of panel d). For systems reported in panels (c) and (d), $P \approx 3$. NCs of panels (a) and (b) have a diameter of about 1.8 nm. NCs of panels (c) and (d) have a diameter of about 2.4 and 3.0 nm, respectively, with core diameter of $\approx 1$ nm  for systems of panel c and $\approx 1.6$ nm for systems of panel d.
    }
    \label{fig:WF_loc}
\end{figure*}


\subsection{Intrinsic band alignment}
\label{Alignment}

As a preliminary step, a simple estimation of the  band-offset properties of a semiconductor-semiconductor junction can be obtained starting from the intrinsic properties of the isolated materials, i.e. by evaluating a simple {\it{intrinsic energy  band  alignment}}  ({\it{IEBA}}) of the corresponding bulk phases. Obviously, when this scheme is adopted, the effects related to the presence of the true interface (for example bonds, strain or defects), are neglected. \\
\noindent
The electronic affinities $\chi$ of crystalline Si and Ge are $\chi_{Si}=4.05$ eV and $\chi_{Ge}=4.00$ eV,  while their  energy gaps are E$_\text{g}^{Si}=1.12$ eV and E$_\text{g}^{Ge}=0.66$ eV, respectively. As a consequence, the band alignment between  Si and Ge bulks leads to an intrinsic type II offset, with the valence band maximum (VBM) localized on Ge (valence band offset, VBO~$\approx$~0.46 eV) and the conduction band minimum (CBM) localized on Si (conduction band offset, CBO~$\approx$~$0.05$ eV). 
A type II offset was also theoretically predicted within DFT and GW schemes  by aligning, with respect to the vacuum level, band-edge energies  of  H-terminated Si and Ge surfaces \cite{Marri_PCCP_2020}. In this case, the CBO (VBO) was estimated to be $0.08$ ($0.59$) eV. A larger CBO is expected when the energy levels alignment is evaluated between isolated Si and Ge nanostructures of the same size.  For instance, the DFT energy levels alignment of two spherical H-terminated Si and Ge NCs of about $2.8$ nm of diameter leads to a CBO of about $0.3$ eV, while  the VBO is reduced to $0.27$ eV. This result is not surprising, because Ge shows a stronger QCE than Si, in the conduction band, upon size reduction \cite{Bostedt_APL_2004,AmatoPRB2009}. Indeed, it has been both theoretically and experimentally proven that, when reducing the size of the system, the bandgap in semiconductors opens with a fixed ratio of the valence and conduction band edge shift \cite{Buuren_PRL_1998,Ru_J_Phys_Cond_Mat_2003,Reboredo_PRB_2000,Beckman_PRB_2006,AmatoPRB2009}, that is $\Delta E_{VBM}^{Si} \div \Delta E_{CBM}^{Si} \approx 2$ for Si and $\Delta E_{VBM}^{Ge} \div \Delta E_{CBM}^{Ge} \approx 1$ for Ge.\\
\noindent
Band-offset properties of low dimensional SiGe  heterostructures, however, cannot be uniquely determined by only considering the intrinsic properties of the Si and Ge materials or by simply aligning the energy levels of isolated nanosystems of similar size~\cite{AmatoCR2014}. They also depend on additional parameters, like for instance the QCE and strain, that are related to the microscopic properties of the interface and, therefore, have to be studied case by case.



\subsection{Si/Ge and Ge/Si CSNCs band-offset properties}
\label{CSNCs}
Results discussed in Sect. \ref{Alignment}, obtained by aligning the energy levels of Si and Ge systems of different dimensionality (bulk, surfaces and NCs of the same size),  predict a type II offset for SiGe heterostructures.
However, when low-dimensional structures are considered, the presence of a Si/Ge interface (not explicitly included, for instance, in the {\it{IEBA}}  model), as well as the different effects that strain and the QCE may have on Si and Ge, can affect the spontaneous formation of a type II offset.
This is particularly true in core-shell nanostructures, where the two materials occupy regions with different shapes and thicknesses. \\
\noindent
In order to investigate the mechanisms that influence the band-offset properties of Si/Ge and Ge/Si CSNCs, we study Si(core)/Ge(shell) and Ge(core)/Si(shell) NCs  with different sizes and compositions.
As a first step, starting from small pristine Si and Ge NCs of nearly 1.8 nm of diameter (Si$_{147}$H$_{100}$ and Ge$_{147}$H$_{100}$, respectively) we generate a set of small Si/Ge and Ge/Si CSNCs  with a different ratio of Si and Ge atoms. 
The Si(core)/Ge(shell) (Ge(core)/Si(shell)) NCs are obtained, starting from spherical Ge (Si) NCs, by replacing the Ge (Si) atoms within an internal sphere, centered in the NC, with Si (Ge) atoms. 
Here we follow the notation Si$_x$Ge$_y$H$_z$ (Ge$_x$Si$_y$H$_z$)  to identify an H$_z$-terminated spherical CSNC constituted by $x$ Si (Ge) atoms located in the core and $y$ Ge (Si) atoms located in the shell region. By changing the radius of the internal sphere, we modify the CSNCs composition. The atomic positions  are then optimized. To characterize these systems, we introduce the parameter $P$, which is defined as the ratio between the number of atoms in the shell, $N_{shell}$, and that in the core, $N_{core}$.\\
\noindent
The HOMO and LUMO states localization for Si/Ge and Ge/Si CSNCs of about 1.8 nm of diameter is depicted in Fig. \ref{fig:WF_loc}, panels (a) and (b). Si(core)/Ge(shell) and Ge(core)/Si(shell) NCs manifestly show a different behavior. As for Si(core)/Ge(shell) NCs,  near-valence-edge states (in particular the HOMO state depicted in Fig. \ref{fig:WF_loc}, panel b) are mainly localized in the shell region, while  near-conduction-edge states (and particularly the LUMO state) are mainly localized in the core region, thus leading to the formation of a type II offset. This result is coherent with the discussion presented in Sect. \ref{Alignment}.
Noticeably, this behavior is independent of the CSNCs composition, i.e. on the parameter $P$. On the contrary, in  Ge(core)/Si(shell) NCs,  both the near-valence and near-conduction-edge states (in particular the HOMO and LUMO states, see Fig. \ref{fig:WF_loc} panel a) are mainly localized in the core region, that is on Ge atoms.  Hence, in this case, we do not observe the formation of a type II junction, as the offset presents a type I character. Even in this case, the results are independent on the CSNCs composition.\\
\noindent
Similar results are obtained for both CSNCs with a diameter of $2.4$ and $3.0$ nm. This is shown in the right panel of Fig. \ref{fig:WF_loc}, where the HOMO and LUMO states localization is depicted for the  Si$_{71}$Ge$_{222}$H$_{172}$  and the Ge$_{71}$Si$_{222}$H$_{172}$ (panel c) and for the Si$_{147}$Ge$_{486}$H$_{300}$  and  Ge$_{147}$Si$_{486}$H$_{300}$  (panel d)  (for simplicity, we report only the HOMO and LUMO wavefunctions localization for CSNCs with P $\approx 3$). In all these cases, Si(core)/Ge(shell) NCs show a type II offset, while the band-offset character of the Ge(core)/Si(shell) NCs mainly resemble that of a type I heterostructure, with both the HOMO and LUMO states mainly localized on the Ge atoms, in particular the LUMO in the outermost part of the core region, near the Si/Ge interface. This is an important point that will be resumed later.
\begin{figure*}[t!]
    \centering
    \includegraphics[width=0.9\textwidth]{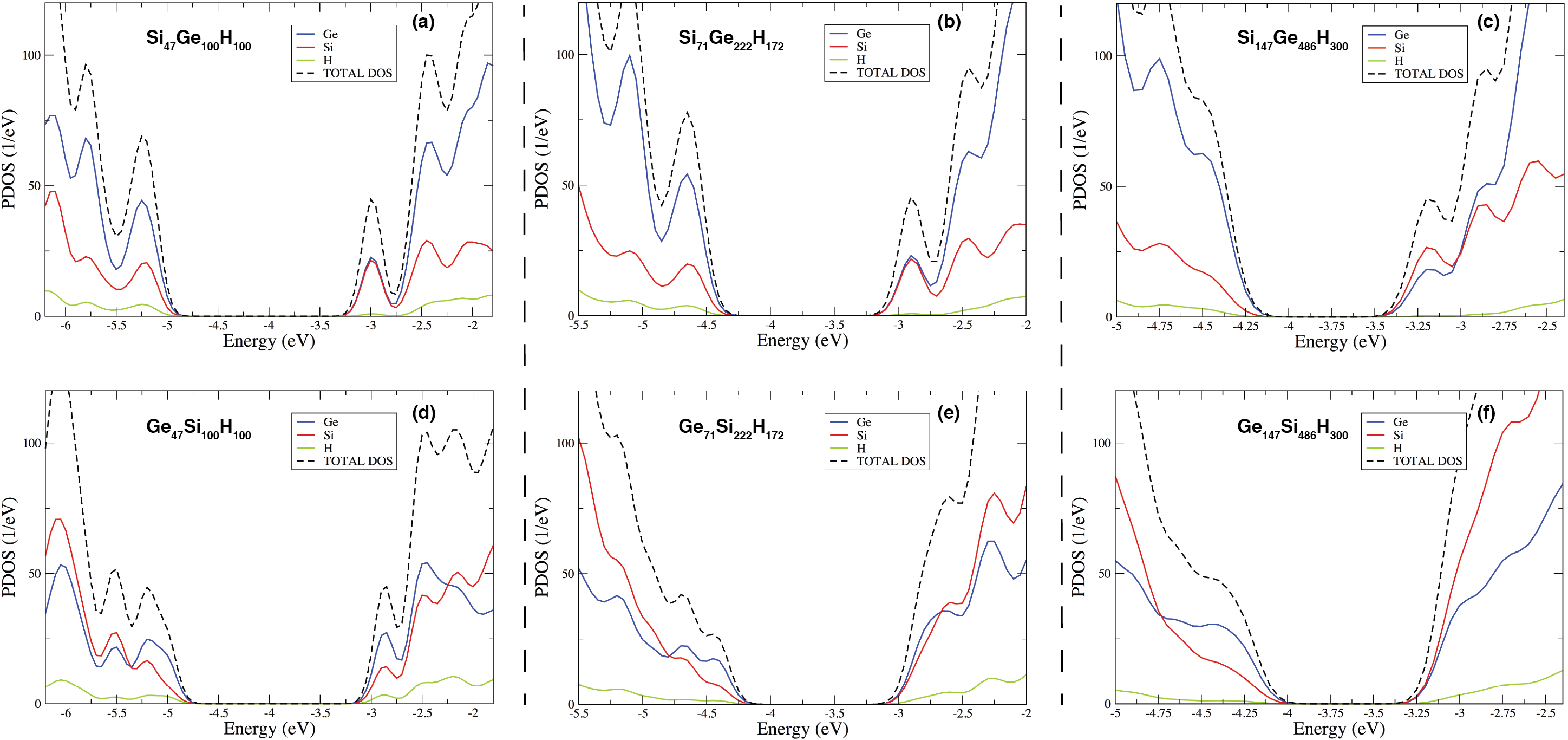}
    \caption{(Color online) Total density of states  (black-dashed line)  and PDOS (colored lines) are reported for Si(core)/Ge(shell) and Ge(core)/Si(shell) NCs of diameters  1.8, 2.4 and 3.0 nm. Contributions from Ge atoms (blue-solid line), Si atoms (red-solid lines) and H atoms (green-solid lines) are underlined in the figure.  
    }
    \label{fig:PDOS}
\end{figure*}
Differences between band-offset characters of Si(core)/Ge(shell) and  Ge(core)/Si(shell) NCs emerge also by  the results of Fig. \ref{fig:PDOS}, where the projected density of states (PDOS) are calculated for the Si$_{47}$Ge$_{100}$H$_{100}$, the Si$_{71}$Ge$_{222}$H$_{172}$, the Si$_{147}$Ge$_{486}$H$_{300}$ (panels a-c, Fig.\ref{fig:PDOS}) and for the Ge$_{47}$Si$_{100}$H$_{100}$, the Ge$_{71}$Si$_{222}$H$_{172}$ and finally  the Ge$_{147}$Si$_{486}$H$_{300}$ (panels d-f, Fig.\ref{fig:PDOS}). For what concerns the Ge(core)/Si(shell) NCs, in particular, electronic states near both the valence and the conduction band-edges have a clear Ge-like character, which contrasts with the formation of a type II offset. This Ge-like character decreases moving from the smallest to the largest NC, which suggests a reduction of the CBO with increasing NC size. 
This behaviour is confirmed by the analysis of the spatial localization of the KS unoccupied states. For what concerns the  Ge$_{47}$Si$_{100}$H$_{100}$, indeed, the first unoccupied state localized on the Si (LUMO$^{Si}$) is $0.29$ eV above the first unoccupied state localized on the Ge (LUMO$^{Ge}$), that in this system constitutes the CBM. The CBO is reduced to $0.24$ eV when the Ge$_{71}$Si$_{222}$H$_{172}$ is considered and moves to only $0.08$ eV for the Ge$_{147}$Si$_{486}$H$_{300}$. Therefore, the Ge$_{147}$Si$_{486}$H$_{300}$ still shows a type I offset but, in this system, DFT predicts LUMO$^{Ge}$ and LUMO$^{Si}$ states to be almost degenerate energy levels, meaning that  Ge(core)/Si(shell) NCs  of $d=3$ nm are close to a type I $\rightarrow$ type II band-offset transition.
\noindent
The calculation of wavefunctions localization  and of the PDOS are useful to understand band-offset properties of both Si/Ge and Ge/Si CSNCs, but does not allow (i) understanding which mechanisms lead to the formation of a well-defined offset, (ii) defining which parameters differentiate the behavior of the Si(core)/Ge(shell) from that of the Ge(core)/Si(shell), and finally (iii) explaining the trend of the band-offset as the size of the NC (and therefore the core and shell thickness) increases. These points will be addressed in the next sections. Since the main differences between the  Si(core)/Ge(shell) and Ge(core)/Si(Shell) are related to the localization of the LUMO state (see Sect. \ref{Alignment}), in the following, we will mainly  focus our attention on the CBO properties.


\subsection{Si/Ge and Ge/Si core shell NCs: the role of quantum confinement}
\label{SiGe_core_shell}
\begin{figure}[h!]
    \centering
    \includegraphics[width=0.45\textwidth]{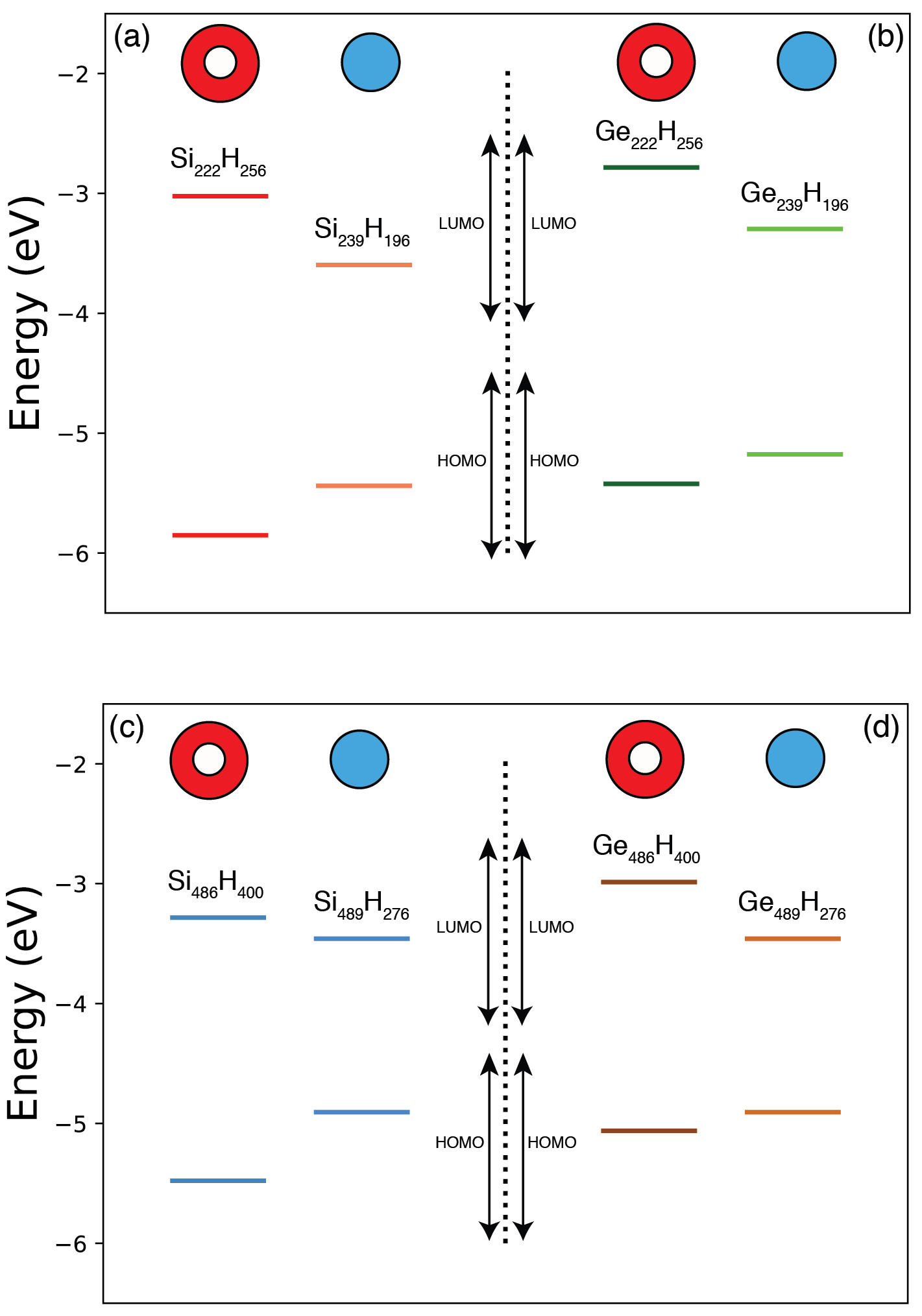}
    \caption{(Color online) The HOMO and LUMO energies calculated for the   nanostructured spherical caps  Si$_{222}$H$_{256}$$_{s}$ and Ge$_{222}$H$_{256}$$_{s}$ (Si$_{486}$H$_{400}$$_{s}$ and Ge$_{486}$H$_{400}$$_{s}$) and for the spherical  Si$_{239}$H$_{196}$ and   Ge$_{239}$H$_{196}$ (Si$_{489}$H$_{276}$  and Ge$_{489}$H$_{276}$) NCs are reported in panels (a) and (b) (panels (c) and (d)). Different colors identify different structures. The energy scale is obtained through the vacuum level alignment rule where the vacuum energy is set to 0 eV.
}
    \label{fig:QCE}
\end{figure}

\noindent
\noindent
To obtain a semi-quantitative analysis of the mechanisms influencing the formation of the band-offset,  we consider two separated systems obtained from both the CSNCs with a diameter of $2.4$ and $3.0$ nm ($P\approx 3$). The first ones are generated by extracting the core region from the CSNCs and then capping its surface with hydrogen atoms. The second ones are obtained by extracting the shell region from the CSNCs and, again, passivating with hydrogen all the internal dangling chemical bonds. As a first step, only additional hydrogen positions are relaxed, keeping both Si and Ge atomic positions unaltered, thus preserving the strain induced by the formation of a Si/Ge interface on both the core and the shell regions.
Since in this case the Si/Ge interface is not explicitly taken into account, the adoption of such a model cannot lead to a precise  quantitative analysis of the band-offset properties of Si/Ge and Ge/Si CSNCs. 
For instance, by impeding the wavefunctions delocalization on both core and shell region, it leads to an overestimation of the confinement of the electronic charge density. However, it is a good approximation  to (i) understand if the QCE has a different relevance in the core and shell regions and (ii) clarify the role played by strain or, more generally, by the structural distortions induced by the formation of a Si/Ge interface.\\
\noindent
The obtained systems are, on one side, the NCs cores Si$_{71}$H$_{84}$$_{c}$, Ge$_{71}$H$_{84}$$_c$, Si$_{147}$H$_{100}$$_{c}$ and Ge$_{147}$H$_{100}$$_c$ (the first two extracted from the CSNCs with $d=2.4$ nm, the other two from the CSNCs with $d=3.0$ nm), and, on the other side, the nanostructured shell caps Si$_{222}$H$_{256}$$_s$, Ge$_{222}$H$_{256}$$_s$, Si$_{486}$H$_{400}$$_s$ and   Ge$_{486}$H$_{400}$$_s$ (same as above). The subscripts $c$ and $s$ indicate that the related (nano)structure is obtained by extracting the core or the shell of the NC, respectively.
The structures are then fully relaxed to evaluate the effects induced by strain on the band-offset properties. The obtained  {\it{unstrained}} systems are identified with the label {\it{relax}}. We refer to  Sect. \ref{strain} for the discussion concerning the role played by strain and we focus here on the connection between the shape of the nanostructures and the QCE, that is on the effects induced by core and shell conformation on the charge density confinement.\\
\noindent
To evaluate the effects of QC,  we  consider   structures  obtained by extracting the shell region from the CSNCs - i.e. the nanostructured  shell caps
Si$_{222}$H$_{256}$$_{s}$,  Ge$_{222}$H$_{256}$$_{s}$,  Si$_{486}$H$_{400}$$_{s}$ and  Ge$_{486}$H$_{400}$$_{s}$ - and we compare their electronic properties with the ones of H-terminated spherical NCs containing a similar number of Si or Ge atoms - i.e. the Si$_{239}$H$_{196}$, the Ge$_{239}$H$_{196}$, the Si$_{489}$H$_{276}$ and the Ge$_{489}$H$_{276}$ NCs. 
The goal is to understand how the QCE depends on the shape of the nanostructures, and especially what happens when we move from a spherical nanostructure (e.g. the core of the CSNC) to a nanostructured cap (e.g. the shell of the CSNC) containing the same number of atoms. Noticeably, this kind of analysis will help us clarify if QC is generally more relevant in the core or in the shell region of CSNCs. \\
\noindent
After calculating the electronic properties, we analyze the energy level alignment with respect to the vacuum level. The obtained results are reported in Fig. \ref{fig:QCE}.
They show that both energy gaps and ionization potential IP (measured as the energy distance between the HOMO and the vacuum  level) are larger in the nanostructured shell caps than in the corresponding spherical NCs, while the electronic affinity EA (measured as the energy distance between the LUMO and the vacuum level) is smaller.
For instance, the EA of the Si$_{222}$H$_{256s}$
is, in absolute value, $0.57$ eV lower than the one of the Si$_{239}$H$_{196}$, while the EA of the Si$_{486}$H$_{400s}$ is $0.18$ eV lower than the one of the Si$_{489}$H$_{276}$. Considering that strain, as we will show later, does not essentially affect the energy of the LUMO$^{\text{Si}}$, we can safely affirm that the QCE is more pronounced when Si atoms are distributed in a shell cap than inside a sphere. Regarding Ge nanostructures, we find that the calculated EA for the Ge$_{222}$H$_{256s}$ is, in module,  $0.51$ eV lower than the one of the Ge$_{239}$H$_{196}$ and that the EA obtained for the Ge$_{486}$H$_{400s}$ is $0.47$ eV lower than the one of the Ge$_{489}$H$_{276}$. Even for Ge, therefore, we can say that the QCE is more pronounced when atoms are distributed in a shell cap than inside a sphere. However, in this case the differences in the calculated EA cannot be ascribed only to the core and shell shape but, as will be later shown, are also affected by strain. Anyway, our conclusions do not change if we repeat the same analysis by considering the fully relaxed shell  nanostructures.\\
\noindent
The obtained results point out that, for reasons related to the geometry of the system, QC acts differently in the core and in the shell regions of CSNCs. Therefore, the band-offset character of Si/Ge and Ge/Si CSNCs cannot be, in principle, determined by simply  considering the band energy alignment of Si and Ge bulks, surfaces or NCs of the same size. \\
\noindent
Starting from these findings, we can try to give a preliminar interpretation of the data of Fig. \ref{fig:WF_loc}, in particular for the Ge(core)/Si(shell) NCs whose band-offset properties deviate from the ones predicted by the {\it{IEBA}} scheme.
In these systems, two different concurrent effects contribute to the formation of the band-offset, and in particular they influence the energy levels alignment of the unoccupied states. The first effect, already discussed in Sect.~\ref{Alignment}, implies that, in similar Si and Ge nanostructures, QC acts more markedly on the  LUMO$^{\text{Ge}}$ than on the LUMO$^{\text{Si}}$. In low-dimensional systems, therefore, the LUMO$^{\text{Ge}}$ is shifted toward higher energies, that is, it tends to get closer to the vacuum level. This effect would strengthen the intrinsic type II offset that generally characterizes Si/Ge junctions.
On the other hand, the electronic charge confinement is generally larger in the shell than in the core region, as shown in Fig.\ref{fig:QCE}.
This effect is in competition with  the one mentioned above and moves the LUMO$^{\text{Si}}$ to higher energies (that is, energies closer to the vacuum level) than the LUMO$^{\text{Ge}}$. 
Despite  the Ge(core)/Si(shell) NCs of Fig. \ref{fig:WF_loc}  have $N_{core} < N_{shell}$,  the second effects cannot be neglected (also because these CSNCs show a thin shell region, approximately less than $0.7-0.8$ nm), fostering the formation of a type I offset.
Obviously, the larger the shell, the weaker QC is in this region.
The two aforementioned effects are instead combined to enhance the shift of the LUMO$^{\text{Ge}}$ toward higher energies when the Si(core)/Ge(shell) NCs are considered. In this case, the different relevance of the QCE in the core and shell regions  strengthens the formation of a type II offset. The quantum confinement of the electronic charge induced by both core and shell morphology is not, however, the only parameter we have to consider. A fundamental  role is also played by strain, especially when small CSNCs are considered. Effects induced by strain on the band-offset properties of the considered systems will be discussed in the next section.

\subsection{SiGe and GeSi core shell NCs: the role of strain}
\label{strain}

\begin{figure*}[t!]
    \centering
    \includegraphics[width=0.9\textwidth]{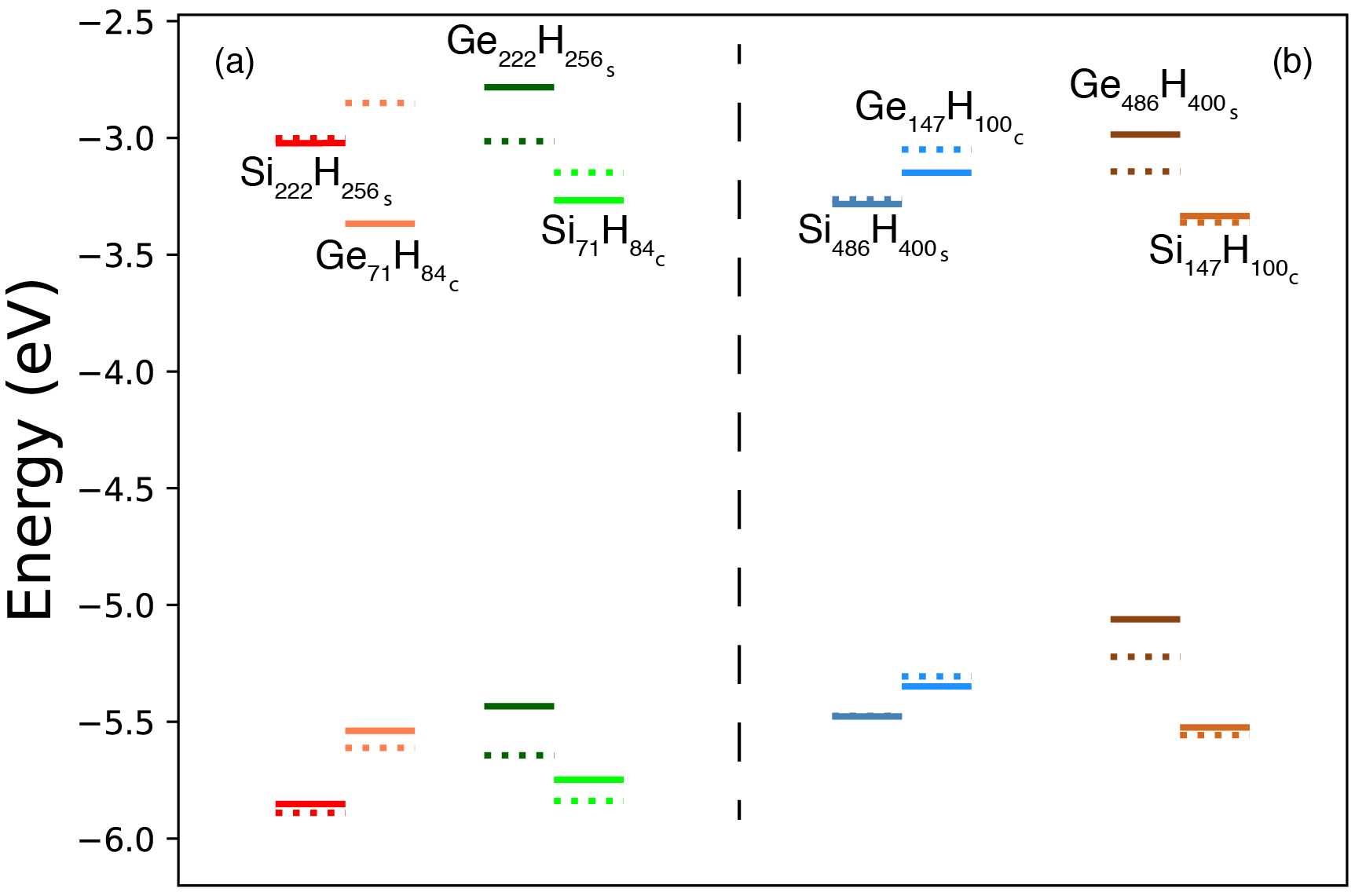}
    \caption{(Color online) In panel (a), solid lines refer to the HOMO and LUMO energies calculated with respect to the vacuum level for the core and shell structures extracted from the CSNCs of $d=2.4$nm, that is the Ge$_{71}$Si$_{222}$H$_{172}$ and the Si$_{71}$Ge$_{222}$H$_{172}$. Dotted lines refer to the relaxed structures. In panel (b) the same analysis is performed considering the systems obtained from the CSNCs of $d=3.0$nm, that is the Ge$_{147}$Si$_{486}$H$_{300}$ and the Si$_{147}$Ge$_{486}$H$_{300}$. The energy scale is obtained through the vacuum level alignment rule where the vacuum energy is set to 0 eV.
    }
    \label{fig:ELABN}
\end{figure*}

\begin{figure}[t!]
    \centering
    \includegraphics[width=0.45\textwidth]{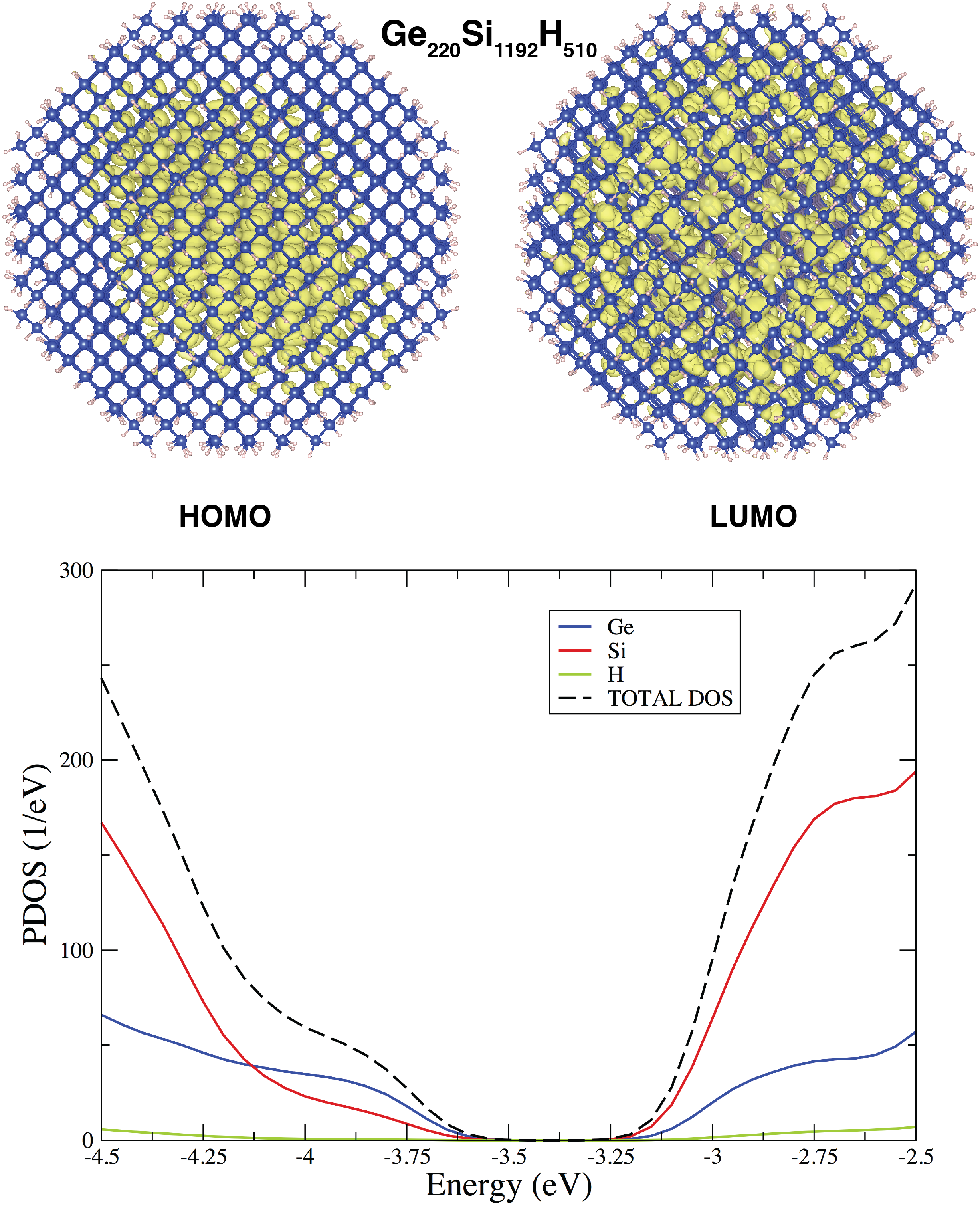}
    \caption{(Color online) In panel (a) we report the localization of the HOMO and LUMO states for the  Ge$_{220}$Si$_{1192}$H$_{510}$ CSNC. In panel (b) the calculated PDOS is reported. Both panels clearly indicate the formation of a type II offset.
    }
    \label{fig:type_2}
\end{figure}

In order to investigate the effects generated on the electronic properties by strain, we focus our attention on the core and shell structures extracted from the CSNCs. These systems can be analyzed to separate and understand the role played by both QC and strain. In  Fig. \ref{fig:ELABN}, solid lines of the panel (a),  we report the energy levels alignment  calculated for the LUMO and HOMO states for the structures extracted from the CSNCs with a diameter of $2.4$ nm. The band diagram is thus obtained for the Si$_{222}$H$_{256s}$  and  the Ge$_{71}$H$_{84c}$, that is the  systems extracted from the $d=2.4$ nm Ge/Si CSNC, and for the Ge$_{222}$H$_{256s}$ and the Si$_{71}$H$_{84c}$ (obtained starting from the $d=2.4$ nm Si/Ge CSNC). In panel (b) the same analysis is performed for the nanostructures extracted from the  CSNCs with a diameter of $3.0$ nm. In both panels (a) and (b), dotted lines refer to the fully relaxed structures. 
A comparison between solid and dotted lines will help to clarify the role played by strain, as it will be discussed later. First, we focus our attention on the results identified by solid lines, in order to discuss the accuracy of a model based on the study of the core and shell regions taken separately.\\
\noindent
Initially, we note that the  band-offsets derived from the energy levels alignment of panel (a) are in agreement with the outcomes of  Fig. \ref{fig:WF_loc} panel (c), that is a type II offset  for the systems extracted from the  Si(core)/Ge(shell) NCs and a type I offset for the structures generated from the Ge(core)/Si(shell) NCs. The CBO obtained by the energy levels alignment of the  Si$_{222}$H$_{256s}$  and  the Ge$_{71}$H$_{84c}$  (Ge$_{222}$H$_{256s}$ and Si$_{71}$H$_{84c}$) is $0.34$ eV ($0.48$ eV), a value that approaches  the one obtained for the corresponding Ge/Si (Si/Ge) CSNC,  that is $0.24$ eV ($0.33$ eV). These results point out a discrepancy of about $0.1 - 0.2$ eV between the CBO obtained from the results of Fig. \ref{fig:ELABN} and the ones corresponding to the structures of Fig. \ref{fig:WF_loc}, panel (c). \\
\noindent
We now move on to consider the results reported in panel (b) of Fig. \ref{fig:ELABN}. 
The energy level alignment of the systems extracted from the $3.0$ nm Si(core)/Ge(shell) NC (that is the Ge$_{486}$H$_{400}$ and the Si$_{147}$H$_{100}$), presents a type II character, in agreement with the results of Fig. \ref{fig:WF_loc}, panel d. Regarding the $3.0$ nm Ge(core)/Si(shell) NC, instead, the model applied in Fig. \ref{fig:ELABN} fails to predict a type I offset, by moving up 
 the crossing  between the LUMO$^{\text{Si}}$ and the LUMO$^{\text{Ge}}$ levels, and thus the transition from the type I to the type II offset. This result is not surprising because, as underlined before, a model based on the energy levels alignment of structures extracted from CSNCs cannot exactly define neither the CBO nor the LUMO$^{\text{Si}}$-LUMO$^{\text{Ge}}$ crossing point. Therefore it cannot accurately predict at which diameter the type I $\rightarrow$ type II transition occurs. As a good approximation, however, it can indicate if we are close to observing such a transition.
Noticeably, the type I offset calculated for the whole Ge$_{147}$Si$_{486}$H$_{300}$ shows a CBO of only $0.08$ eV, that is the LUMO$^{\text{Ge}}$ and the LUMO$^{\text{Si}}$ are almost degenerate, while the HOMO - LUMO energy alignment obtained for the separated Si$_{486}$H$_{400s}$ and  Ge$_{147}$H$_{100c}$ partes point out a type II offset with a CBO of only $0.13$ eV, a result that confirms the accuracy of the method with an error in the estimation of the band offset properties of  about $0.2$ eV. Starting from these considerations, we compare the solid and dotted lines of Fig. \ref{fig:ELABN}. As a result, we can observe that distortions induced by the formation of a Si/Ge interface mainly affect the band-offset properties of the smaller structures, that is the ones reported in panel (a), which refer to the CSNCs with a diameter of $2.4$ nm and, above all, they impact on the energy of the LUMO$^{Ge}$ state. 
In particular, by focusing on the results reported in panel (a) of Fig. \ref{fig:ELABN}, we observe an increase of about $0.51$ eV of the LUMO$^{\text{Ge}}$ energy when we move from the Ge$_{71}$H$_{84c}$ (solid orange line) to the Ge$_{71}$H$_{84c}^{relax}$ (dotted orange line) and a lowering of 
$0.23$ eV of the energy of the  LUMO$^{\text{Ge}}$ level when we move from the Ge$_{222}$H$_{256s}$ to the Ge$_{222}$H$_{256s}^{relax}$ (from solid to dotted green lines of panel a). As a consequence, structural distortions induced by the formation of a Si/Ge interface strengthen the type I offset in the $d=2.4$ nm Ge(core)/Si(shell) NC and the type II offset in the $d=2.4$ nm Si(core)/Ge(shell) NC. In order to understand these changes, we have to remember that, as a consequence of strain, in both the Si(core)/Ge(shell) and the Ge(core)/Si(shell) NCs, the LUMO$^{Ge}$ state is mainly localized in the proximity of the Si/Ge interface. This characteristic is also present in the isolated core and shell nanostructures. In particular, the LUMO$^{Ge}$ is localized in the outermost part (that is, near the surface) of the Ge$_{71}$H$_{84c}$ and  in the proximity of the internal surface when the nanostructured spherical cap Ge$_{222}$H$_{256s}$ is taken into account. When systems are fully relaxed, and the distortions induced by the formation of a Si/Ge interface are removed, the LUMO$^{Ge}$ state appears to be more localized in the central part of the nanostructures. This implies a more pronounced confinement of the LUMO$^{Ge}$ state in the structure extracted from the core, with a consequent shift to higher energies of the LUMO$^{Ge}$  when strain is removed (from solid to dotted orange lines). At the same time, a weaker confinement of the  LUMO$^{Ge}$ state in the  structure extracted from the shell, with a consequent reduction of the  LUMO$^{Ge}$ energy (from solid to dotted green lines) is found.
The same trends can be observed for the Ge structures of panel (b) but, in this case, due to the larger size of the systems, the effects induced by strain are less relevant. Regarding Si nanostructures, we do not observe drastic changes in the HOMO and LUMO positions when strain is removed (changes are always less than $0.1-0.15$ eV), confirming that Si NCs are less sensitive to the strain than Ge ones \cite{Weissker_PRB_2003}.\\
\noindent 
The results of Fig.\ref{fig:ELABN} clarify that a second condition has to be verified in order to observe a type II offset in the Ge(core)/Si(shell) NCs, that is, the core has to be sufficiently large to reduce the effects induced by strain. When this last condition is not verified, structural distortions present in the core region (induced by the Si and Ge lattice mismatch) induce a lowering of the LUMO$^{Ge}$ energy, strengthening the type I offset.
As a consequence, starting from the $d=3$ nm Ge$_{147}$Si$_{486}$H$_{300}$ NC, which shows a type I offset, we cannot obtain a type II heterostructure by simply increasing the shell thickness while keeping the size of the NC constant. By doing so, indeed, we also reduce the core extension, thus increasing the relevance of strain in this region. The result would be a simultaneous lowering of both the LUMO$^{Si}$ and  LUMO$^{Ge}$ energies, the former induced by the slight reduction of the QCE in the shell region, the latter by increasing strain in the core region, which would not produce changes in the band-offset character. To verify this point, we have performed calculations considering the $d=3$ nm  Ge$_{71}$Si$_{562}$H$_{300}$ (core diameter of about $1.0$ nm, shell thickness of about $1.0$ nm) still obtaining a type I offset with a CBO even increased to $0.13$ eV.\\
\noindent The diameter $d=3$ nm represents therefore a sort of critical size for the Ge(core)/Si(shell) NCs. Below this threshold, we clearly observe a type I offset with both the HOMO and LUMO states mainly localized on the Ge atoms; the same offset is observed  for Ge(core)/Si(shell) NCs with a diameter of about $3$ nm, but in this case we are in proximity of the type I $\rightarrow$ type II transition. Finally, a type II offset can result  for Ge(core)/Si(shell) NCs with $d>3$ nm. This condition  guarantees the possibility of obtaining CSNCs with, at the same time, a sufficiently large shell to reduce the QCE in the Si region, and a sufficiently large core to reduce the effects induced by strain in the Ge region. As an example, we report in Fig. \ref{fig:type_2} the case concerning the  
 Ge$_{220}$Si$_{1192}$H$_{510}$ ($d \approx 4$ nm, $d_{core} \approx 2$ nm) which shows a type II band-offset, with the HOMO mainly localized in the core around the Ge atoms, and the LUMO localized on the Si atoms, thus outside of the core region, in between the Si/Ge interface and the outermost part of the shell. The formation of a type II band-offset is also confirmed by the calculated PDOS, as reported in Fig. \ref{fig:type_2} bottom panel.

\section{Conclusions}
\label{conclusions}
In this paper, Density Functional Theory has been adopted in order to investigate the mechanisms involved in the formation of the band-offset of Si/Ge and Ge/Si CSNCs, and in particular to discern the role played by the QCE and strain. NCs with diameters ranging from $1.8$ to $4.0$ nm have been considered. This analysis is crucial because, depending on the band-offset character - type I or type II -, NCs are more suitable to be engineered in  light-emitting (type I) or in photovoltaic (type II) devices. Our results point out that both QC and strain contribute to the formation of type II offset in Si(core)/Ge(shell), with the HOMO localized in the Ge shell and the LUMO localized in the Si core. In these systems, therefore, the band-edge properties resemble those obtained by simply considering the intrinsic properties of the Si and Ge materials. The analysis is far more subtle for Ge(core)/Si(shell) NCs. In these NCs the HOMO state is always localized in the core region while the LUMO localization depends on the geometry of the system, which determines the relevance of QC and strain on both core and shell regions. Our calculations point out that QC is generally  more pronounced in the shell than in the core region. In Ge(core)/Si(shell) NCs, indeed, the QC contributes to moving the LUMO$^{Si}$ level to higher energies. On the contrary, the strain induced by the formation of a Si/Ge interface mainly affects the LUMO$^{Ge}$ state, lowering its energy. As a result, in Ge/Si CSNCs with a diameter less than $3$ nm, the LUMO is always localized on the Ge inducing a type I offset. Our outcomes point out that, depending on the band-offset properties, spherical GeSi CSNCs can be grouped into three different classes. CSNCs with a diameter less than $3$ nm are clearly characterized by a type I offset. CSNCs with a diameter of about $d=3$ nm still show a type I offset though they have the critical size for observing a type I $\rightarrow$ type II transition. Finally, CSNCs with  diameters above $d=3$ nm can present a type II offset because, in this case, core and a shell are sufficiently large to reduce the effects induced by strain in the internal core region, and by the QCE, in the external shell. This has been directly demonstrated in the case of a large Ge$_{220}$Si$_{1192}$H$_{510}$ CSNC, where calculations reveal indeed a type II band-offset.

\section{Appendix}
\label{appendix}

\begin{table*}[t!]
\centering
\caption{DFT-LDA energy gaps calculated for Ge(core)/Si(shell) and Si(core)/Ge(shell) of different compositions and size are reported in the table. As a reference, for the smaller NCs, the pristine Ge$_{147}$H$_{100}$ and Si$_{147}$H$_{100}$ show an energy gap of 2.25 and 2.19 eV, respectively. For what concerns the Ge$_{35}$Si$_{122}$H$_{100}$, we have also calculated the  GW energy gap which settles to 3.5 eV.}
\label{SiGe_gaps}
\begin{tabular}{c c c c c}
 \hline
  \multicolumn{1}{|c|}{\bf diameter (nm)} &  
 \multicolumn{1}{|c|}{\bf Ge(core)/Si(shell)} &  
\multicolumn{1}{|c|}{\bf Energy Gap (eV)} & 
\multicolumn{1}{|c|}{\bf Si(core)/Ge(shell)} &  
\multicolumn{1}{|c|}{\bf Energy Gap (eV)}  \\
 \hline
 1.8 & Ge$_{17}$Si$_{130}$H$_{100}$ &  2.09 &  Si$_{17}$Ge$_{130}$H$_{100}$ & 2.10 \\
 1.8 &  Ge$_{35}$Si$_{122}$H$_{100}$ & 2.12 &   Si$_{35}$Ge$_{122}$H$_{100}$ & 2.12 \\
 1.8 &   Ge$_{47}$Si$_{100}$H$_{100}$ & 2.11 &   Si$_{47}$Ge$_{100}$H$_{100}$ & 2.12 \\
 1.8 &   Ge$_{71}$Si$_{76}$H$_{100}$ & 2.16 &   Si$_{71}$Ge$_{76}$H$_{100}$ &  2.10 \\
 \hline
 2.4 & Ge$_{71}$Si$_{222}$H$_{1772}$ &  1.54 &  Si$_{71}$Ge$_{222}$H$_{172}$ & 1.67 \\
 \hline
 3.0 & Ge$_{147}$Si$_{486}$H$_{300}$ &  1.19 &  Si$_{147}$Ge$_{486}$H$_{300}$ & 1.19 \\
 3.0 & Ge$_{71}$Si$_{562}$H$_{300}$ &  1.17 &  -   & - \\
 \hline
 4.0 & Ge$_{220}$Si$_{1192}$H$_{510}$ &  0.83 &  -   & -  \\
 \hline
\end{tabular}
\end{table*}

\begin{figure}[h!]
    \centering
    \includegraphics[width=0.45\textwidth]{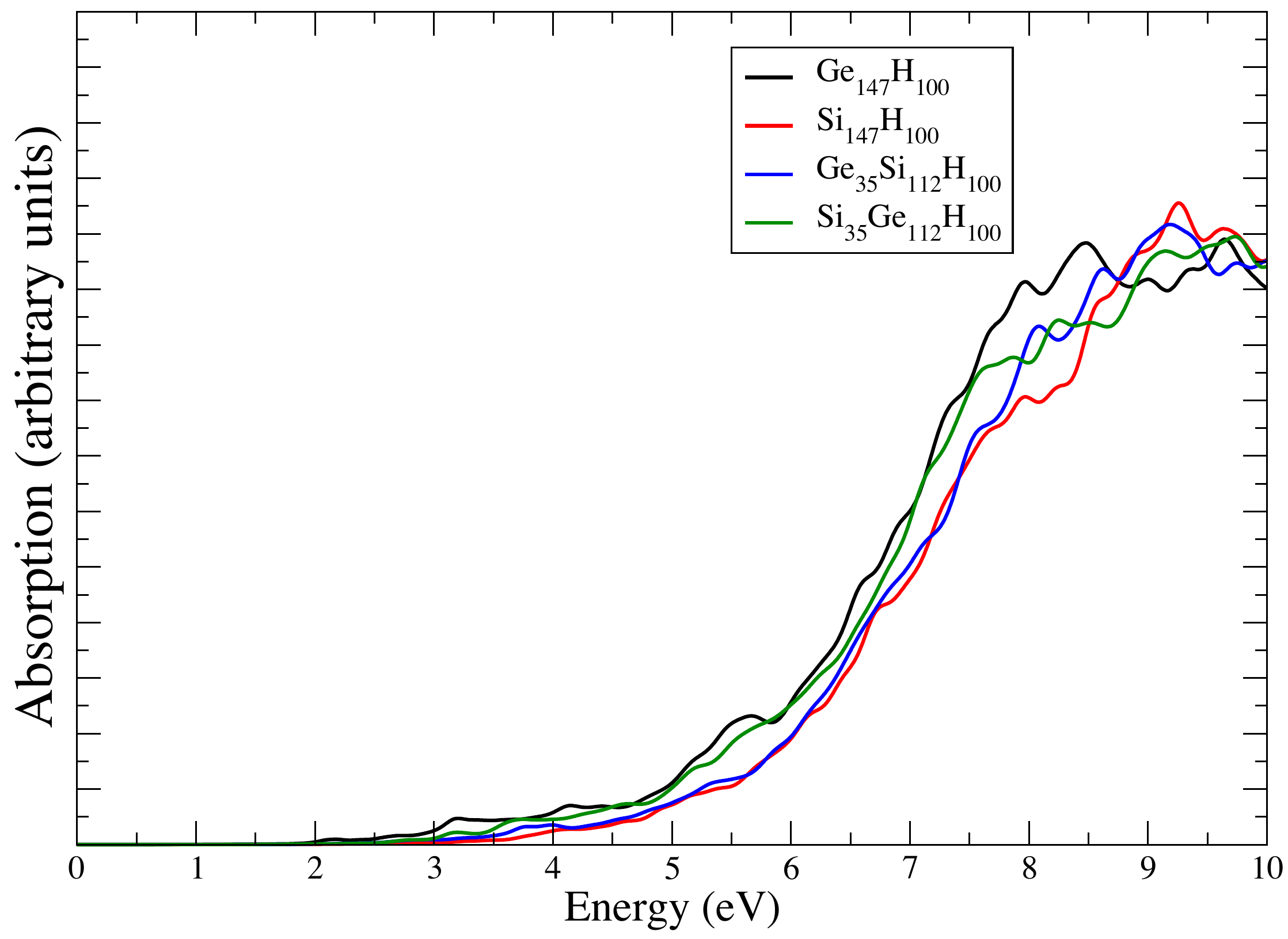}
    \caption{(Color online)  Absorption spectra, calculated for the Ge$_{35}$Si$_{122}$H$_{100}$ and the Si$_{35}$Ge$_{122}$H$_{100}$ NCs are reported in the figure and compared with the ones obtained for the Si$_{147}$H$_{100}$ and the  Ge$_{147}$H$_{100}$.
    }
    \label{fig:Spettro}
\end{figure}

In this section, we report additional information concerning electronic and optical properties of the systems analysed in the manuscript.
Calculated LDA energy gaps $E_{gap}^{LDA}$ are reported in Table \ref{SiGe_gaps}. It is evident that the $E_{gap}^{LDA}$ depend only marginally on the CSNCs composition and, following the typical trend imposed by the QCE, decrease when the NCs size increases. Moreover, by focusing on the smaller NCs, we can also observe that the $E_{gap}^{LDA}$ calculated for the CSNCs do not  strongly differ from those of the pristine Si$_{147}$H$_{100}$ and  Ge$_{147}$H$_{100}$ NCs.\\
\noindent
For the smaller nanocrystals, we have also calculated the absorption spectra using the  Liouville–Lanczos approach to Time-Dependent Density Functional Perturbation Theory, as implemented in the TDDFT tool of the QE package \cite{MALCIOGLU_Comp_Phts_Comm_2011}. The results obtained for the pristine Si$_{147}$H$_{100}$ and  Ge$_{147}$H$_{100}$ NCs and the  Ge$_{35}$Si$_{122}$H$_{100}$ and the Si$_{35}$Ge$_{122}$H$_{100}$ CSNCs are depicted in Fig. \ref{fig:Spettro}.
The spectra calculated for the CSNCs, and their  absorption energy thresholds, fall in-between the ones obtained for the Si$_{147}$H$_{100}$ and the  Ge$_{147}$H$_{100}$ NCs.

\noindent

\begin{figure}[h!]
    \centering
    \includegraphics[width=0.45\textwidth]{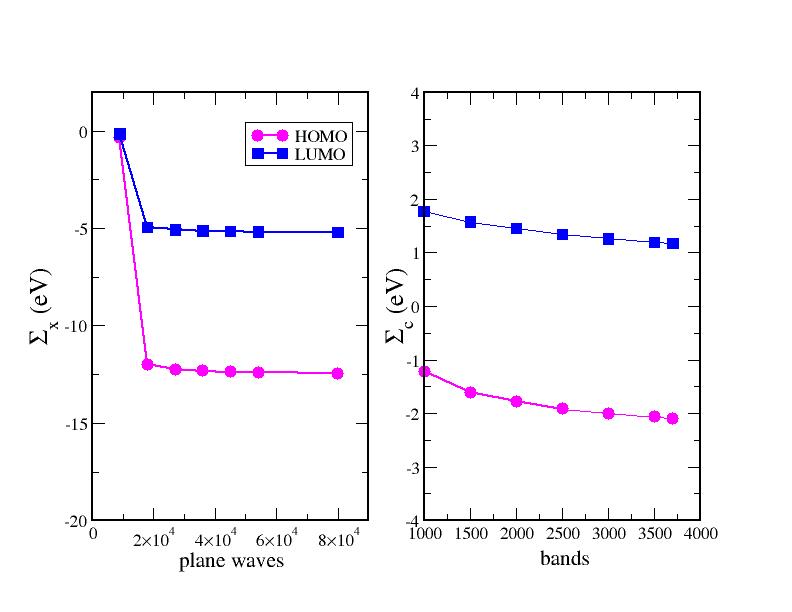}
    \caption{(Color online) Convergence tests for the Self Energy $\Sigma=\Sigma_x+\Sigma_c$ calculated for the HOMO and LUMO levels of the Ge$_{35}$Si$_{122}$H$_{100}$ NC. The correlation part of the self-energy $\Sigma_c$ was evaluated using 30000 plane waves.}
    \label{fig:GW}
\end{figure}

\noindent
In Table \ref{SiGe_gaps}, we also report the GW gap of one of the smallest NCs, i.e. Ge$_{35}$Si$_{122}$H$_{100}$. Quasi-particle calculations were introduced in order to check if a different energy levels alignment could arise when many-body effects are taken into account. Convergence tests are shown in Fig.\ref{fig:GW}. GW corrections are sizeable, opening the DFT gap by about 1.4 eV; nevertheless, no change in the band-edge ordering is observed. In both DFT and GW, the states around the gap are mainly localized on the Ge core, hence giving a type I heterostructure. The first unoccupied state mainly localized on Si is at 0.29 eV above the LUMO in DFT (0.2 eV above the LUMO in GW); analogously, the first occupied Si state is located 0.46 eV below the HOMO (1.0 eV below the HOMO  in GW). Hence, the DFT alignment is confirmed also by quasi-particle calculations.   

\section{Acknowledgments}
\label{Acknowledgements}
I.M., S.G., O.P.  and S. O. thank the Super-Computing Interuniversity Consortium CINECA for support and high-performance computing resources  under the Italian Super-Computing Resource Allocation (ISCRA) initiative, and under PRACE. 
 S.O.  acknowledges support/funding from University of Modena and Reggio Emilia under project ”FAR2017INTERDISC”.
O.P.  and S.G. acknowledge financial funding from the EU MSCA-RISE project  DiSeTCom  (GA 823728) and INFN project TIME2QUEST. Technical support by Dr. Ihor Kupchak is gratefully acknowledged.  M.A. greatly acknowledges the ANR AMPHORE project (ANR-21-CE09-0007) of the French Agence Nationale de la Recherche.


\newpage
\bibliographystyle{aip}
\bibliography{paper}


\end{document}